\documentclass[reprint,prl,superscriptaddress]{revtex4-1}
\usepackage{graphicx}
 \usepackage{amsmath}
\usepackage{bm}

\begin{document}

\title{Localization optoacoustic tomography}
\author{X. Lu\'is De\'an-Ben}
\email{Corresponding author: xl.deanben@helmholtz-muenchen.de}
\affiliation{Institute for Biological and Medical Imaging (IBMI), Helmholtz Zentrum M\"unchen, Neuherberg, Germany}
\author{Daniel Razansky}
\affiliation{Institute for Biological and Medical Imaging (IBMI), Helmholtz Zentrum M\"unchen, Neuherberg, Germany}
\affiliation{Faculty of Medicine, Technische Universit\"at M\"unchen, Munich, Germany}

\date{\today}
\begin{abstract}

Localization-based imaging has revolutionized fluorescence optical microscopy and has also enabled unprecedented ultrasound images of microvascular structures in deep tissues. Herein, we introduce a new concept of localization optoacoustic tomography (LOAT) that employs rapid sequential acquisition of three-dimensional optoacoustic images from flowing absorbing particles. We show that the new method enables breaking through the spatial resolution barrier of acoustic diffraction while further enhancing the visibility of structures under limited-view tomographic conditions. Given the intrinsic sensitivity of optoacoustics to multiple hemodynamic and oxygenation parameters, LOAT may enable new level of performance in studying functional and anatomical alterations of microcirculation.
\newline

\end{abstract}

\maketitle 

Diffraction causes blurring of image features and has been traditionally associated with the spatial resolution limit in light microscopy and other imaging modalities \cite{abbe1873beitrage}. Image resolution is defined as the smallest distance between two points that can be differentiated unambiguously. The resolution of an imaging system is quantified via its point spread function (PSF), corresponding to an image acquired from a point source. It is generally assumed that only points separated by a distance larger than the full width at half maximum (FWHM) of the PSF can be resolved. The FWHM of the PSF may erroneously be interpreted as an "accuracy limit" affecting any dimensional measurement in the images. However, the precision in determining the position of an isolated source can greatly surpass the diffraction limit. Then, if individual sources can be isolated in a certain way, an image can be built by superimposing their estimated positions \cite{betzig1995proposed}. This idea has been successfully exploited in super-resolution fluorescence microscopy methods based on single-molecule localization, independently termed as stochastic optical reconstruction microscopy (STORM) \cite{rust2006sub}, photo-activated localization microscopy (PALM) \cite{betzig2006imaging} or fluorescent photo-activation localization microscopy (FPALM) \cite{hess2006ultra}. Localization microscopy has further been extended to three-dimensional imaging \cite{small2014fluorophore} and recent developments have enabled a never-seen-before resolution in a few nanometers range \cite{balzarotti2016nanometer}. Localization has also been used for overcoming the acoustic diffraction barrier in ultrasound (US) imaging. In this case, super-resolution is achieved by determining the position of individual microbubbles or nanodroplets \cite{viessmann2013acoustic,christensen2015vivo,luke2016super}. Since such particles are moving with the blood flow, they can be localized in different positions in a sequence of B-mode images. An image of vascular structures can then be rendered by superimposing all localized sources. Ultrafast US localization has provided unprecedented images of cerebral microvessels in rodents and has further enabled the characterization of microvascular flow \cite{errico2015ultrafast}.

Being based on light excitation and ultrasound detection, the resolution of optoacoustic (OA) imaging is affected by both optical and acoustic diffraction, typically scaling with 1/200 of the imaging depth \cite{wang2012photoacoustic}. Optical-resolution OA images can be acquired by focusing the excitation light up to a depth of $\sim1\;\mathrm{mm}$ within scattering biological tissues, where super-resolution OA methods similar to those used in fluorescence microscopy have been employed to overcome optical diffraction \cite{danielli2014label,yao2014photoimprint,lee2015vivo}. At deeper regions, OA tomography (OAT) relies on acoustic inversion methods with acoustic diffraction representing an actual resolution barrier. State-of-the art OA systems based on parallel acquisition of signals for each laser pulse with ultrasound arrays have enabled two- and three-dimensional imaging at unprecedented volumetric rates \cite{dean2016functional}. The resolution of such systems has been enhanced via processing a sequence of images acquired by slightly shifting the detection array with a pixel super-resolution approach \cite{he2016improving}. Also, dynamic imaging of flowing individual absorbers has been showcased \cite{dean2017dynamic}, which was also exploited for resolution enhancement with super-resolution optical fluctuations imaging (SOFI) \cite{chaigne2017super}. The increase in resolution achieved with these methods relies on processing aspects such as regularization parameters or the artefacts induced in high order cumulants and is still limited.

\begin{figure}[h!]
 \centering
 \includegraphics[width=\columnwidth]{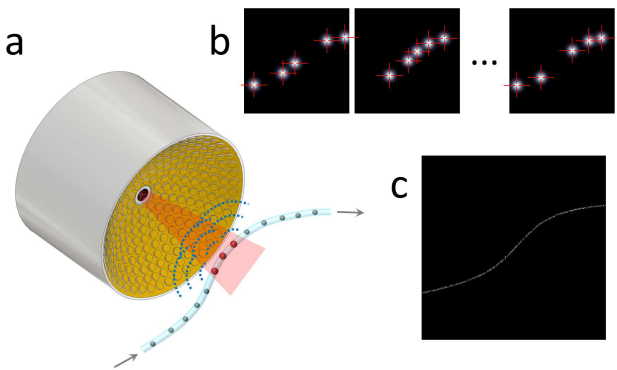}
 \caption{Imaging principle of localization optoacoustic tomography. (a) A spherical array of ultrasound transducers is used to acquire a three-dimensional optoacoustic image of flowing absorbers for each laser pulse. (b) The positions of sparsely distributed absorbers are measured (localized) in a sequence of images. (c) An image is formed by superimposing the localized positions.}
 \label{fig1}
\end{figure}

In this letter, we introduce localization optoacoustic tomography (LOAT) aimed at overcoming the spatial resolution barrier of acoustic diffraction in optoacoustic imaging and tomography. The general concept consists of rapid acquisition of a sequence of three-dimensional OA images from flowing absorbers (Fig.\;\ref{fig1}a). In this way, any structure supporting the particle flow can be accurately mapped by resolving multiple individual absorbers occupying its volume. Provided the absorbers are separated by a distance larger than the diffraction-limited resolution, their individual locations can be accurately determined in each frame of the sequence (Fig.\;\ref{fig1}b). Then, an image can be formed by superimposing a set of points corresponding to the localized positions of the absorbers (Fig.\;\ref{fig1}c).

Experimental validation of the LOAT concept has been performed using a real-time volumetric OAT system \cite{dean2013portable}. It employs a spherical matrix detection array uniformly populated by 256 individual piezocomposite elements having $4\;\mathrm{MHz}$ central frequency and $\sim100\%$ bandwidth in receive mode. The spherical aperture has a radius of curvature of $40\;\mathrm{mm}$ and covers an angle of $90^\circ$. The volume between the imaged sample and the matrix array was filled with agar, which served as an acoustic coupling medium and was further used to hold the imaged samples. Short-duration ($<10\;\mathrm{ns}$) pulses at $720\;\mathrm{nm}$ wavelength, generated by an optical parametric oscillator (OPO) laser (Innolas Laser GmbH, Krailling, Germany), were used for the excitation of OA responses. The matrix array has a central opening through which the light beam was guided using a custom-made fiber bundle (Ceramoptec GmbH, Bonn, Germany). The detected 256 time-resolved OA signals were simultaneously digitized at 40 megasamples per second with a custom-made data acquisition system (Falkenstein Mikrosysteme, GmbH, Taufkirchen, Germany). The acquired raw OA signals were band-pass filtered and deconvolved with the impulse response of the transducer. Image reconstruction was performed with a three-dimensional model-based algorithm implemented on a graphics processing unit (GPU) \cite{ding2017efficient}. Localization of individual absorbers was performed by determining the position of local maxima in the optoacoustic images.

\begin{figure}[h!] 
 \centering
 \includegraphics[width=\columnwidth]{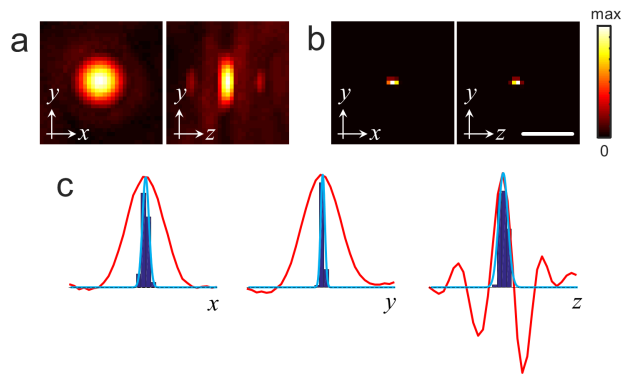}
 \caption{Localization accuracy. (a) Maximum intensity projections of the three-dimensional optoacoustic image of a $30\;\mu\mathrm{m}$ absorbing microsphere. (b) Equivalent image obtained as the three-dimensional histogram of the localized positions in a sequence of 5000 frames. (c) Normalized histograms of the localized positions in the three Cartesian coordinates (fitted Gaussian curves are shown in blue) along with the corresponding profiles (red curves) of the optoacoustic image in (a). Scalebar $400\;\mu\mathrm{m}$.}
 \label{fig2}
\end{figure}

In a first experiment, a single absorbing $~30\;\mu\mathrm{m}$ diameter polyethylene microsphere (Cospheric BKPMS 27-32) was embedded in the agar matrix at approximately the center of the spherical array and a sequence 5000 frames was acquired. The light fluence at the particle location was approximately $10\;\mathrm{mJ}/\mathrm{cm}^2$. The cut-off frequencies of the band-pass filter were set to $0.1$ and $8\;\mathrm{MHz}$ and reconstruction was subsequently performed in a Cartesian grid of $31\times31\times31\;\mathrm{voxels}$ ($1\times1\times1\;\mathrm{mm}^3$). Due to its small size, the microsphere acts as a point absorber, thus the reconstructed volumetric image (Fig.\;\ref{fig2}a) corresponds approximately to the point-spread-function of the OA imaging system. The volumetric image built as the histogram of the localized positions of the absorber for all frames of the sequence is displayed in Fig.\;\ref{fig2}b. The histograms of the $x$, $y$ and $z$ coordinates of the localized positions are shown in Fig.\;\ref{fig2}c along with fitted Gaussian curves. The line profiles through the OA image along the three Cartesian coordinates are also displayed in Fig.\;\ref{fig2}c. In fluorescence microscopy, the localization precision is usually reported as the standard deviation of multiple localized positions of the same source \cite{huang2009super}. Hence, the comparison in Fig.\;\ref{fig2}c illustrates the enhancement in resolution achieved with LOAT as compared with standard OAT. It is also important to note that, while negative values are present in the individual OA image frames (Fig.\;\ref{fig2}c, right) due to the limited-view tomographic geometry \cite{dean2016link}, no negative values are generated in the images formed by localizing individual absorbers.

\begin{figure}[h!]
 \centering
 \includegraphics[width=\columnwidth]{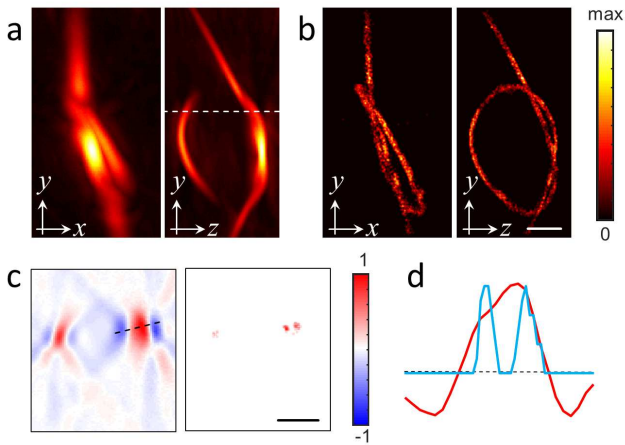}
 \caption{Resolution enhancement in localization optoacoustic tomography. (a) Maximum intensity projections of the three-dimensional optoacoustic image of a $\sim220\;\mu\mathrm{m}$ diameter pipette tip bent to form a knot and filled with ink. (b) Equivalent images obtained by localizing the positions of 3600 flowing $30\;\mu\mathrm{m}$ absorbing microspheres. (c) Comparison of the cross-sections marked in (a) for the standard optoacoustic image (left) and the localization optoacoustic image (right). (d) Comparison of the profiles marked in (c) for the standard optoacoustic image (red) and the localization optoacoustic image (blue). Scalebar $1\;\mathrm{mm}$.}
 \label{fig3}
\end{figure}

In a second experiment, the particles were suspended in ethanol and circulated through a $20\;\mu\mathrm{l}$ Eppendorf microloader pipette tip ($\sim220\;\mu\mathrm{m}$ inner diameter) bent to form a knot. A sequence of 1000 images was acquired, whereas the laser was running at a pulse repetition frequency of $10\;\mathrm{Hz}$. Subsequently, the pipette tip was filled with India ink (Higgins, Chartpak, optical density 20). The acquired signals were band-pass filter between $0.1$ and $3\;\mathrm{MHz}$ in order to emphasize the super-resolution property of the LOAT method by deliberately reducing the effective diffraction-limited spatial resolution of the imaging system. Reconstruction was performed in a Cartesian grid of $320\times320\times320\;\mathrm{voxels}$ ($8\times8\times4\;\mathrm{mm}^3$). The images obtained with standard OAT for the pipette tip filled with ink and those obtained with LOAT are displayed in Figs.\;\ref{fig3}a and \;\ref{fig3}b with their normalized cross-sections and one-dimensional profiles shown in Fig.\;\ref{fig3}c, Fig.\;\ref{fig3}d. The LOAT image was formed by considering 3600 points. Specifically, a binary image was formed by setting those voxels for which at least one particle was localized to 1 and 0 otherwise. A low pass (average) filter with kernel size $3\times3\times3$ was eventually applied to smooth the image. Rotational views of both images are provided in Supplementary Video 1. The actual formation of the super-resolution LOAT image from multiple localized positions of the absorbers is further illustrated in Supplementary Video 2. While LOAT clearly resolves the shape of the knot, this is not possible with the standard diffraction-limited OAT. Apart from attaining significantly better resolution beyond the diffraction limit, LOAT is also responsible for the enhanced visibility of the lateral sides of the pipette tip. The lateral sides are obscured in the regular OAT images due to the limited tomographic view of the matrix detection array \cite{dean2016link}. Thus, superimposition of multiple localized positions aids in mitigating limited-view artefacts and eliminating negative values from the images. 

The presented results demonstrate that LOAT can beat acoustic diffraction and achieve super-resolution OA imaging in three dimensions. Optoacoustics is known for its powerful capability to attain optical absorption contrast at centimeter-scale depths with a much higher resolution than diffuse optical imaging modalities \cite{ntziachristos2010going}. LOAT can further enhance the resolution and enable deep-tissue imaging with unprecedented quality. In the suggested implementation, LOAT is especially suited for imaging vascular structures that can support the flow of extrinsically-administered absorbers. Angiographic imaging is of interest to study cancer, brain function, peripheral vascular diseases and many other conditions and LOAT can play an important role in the diagnosis, treatment monitoring and fundamental understanding of the mechanisms of alteration in the microcirculation. Another important advantage of LOAT is that the rendered images are not affected by lack of visibility under limited-view conditions when tomographic coverage of at least $180^\circ$ around the imaged area cannot be provided. Limited-view acquisitions are common in OA imaging systems and unavoidable if the imaged sample is only accessible from a limited angular range. Much like other methods based on acquisition of a sequence of images of distributed absorbers \cite{dean2017dynamic,gateau2013improving}, LOAT enables visualizing vascular structures in arbitrary directions. Yet, optimal performance of LOAT in living organisms is tightly linked to the development of biocompatible particles capable of generating sufficiently strong signals to be recognized in the presence of a highly-absorbing blood background. Specifically, the dynamic range for ultrasound detection signals must be sufficiently large to cover both the signals generated by red blood cells within the diffraction-limited resolution volume and the signal corresponding to the absorber to be localized. Localization can be further facilitated if the absorber can be differentiated from blood e.g. via spectral or temporal unmixing \cite{dean2017advanced}.

The resolution of LOAT for angiographic applications is ultimately limited by the separation between the smallest capillaries. While the resolution limit in localization fluorescence microscopy is generally associated with the number of photons that a molecule can emit before bleaching occurs \cite{balzarotti2016nanometer}, the LOAT approach exploits a much more versatile optical absorption contrast for which highly photostable agents with long circulation time exist \cite{dean2017advanced,weber2016contrast}. Thus, the resolution in vivo is expected to be determined by the OA generation efficiency of the particular absorbing particle employed, the local light fluence and the number of localized points in the time window of interest. The digital sampling rate of the detected OA response is also important as it affects the maximum spatial frequency achievable in the reconstructed images. Furthermore, it is important to take into account that the PSF of the OA imaging system can be distorted due to speed of sound heterogeneities and acoustic attenuation \cite{dean2011effects,dean2014effects}, and hence also affect the achievable resolution. 

The main limitation of LOAT is the time required to form an image. Although the temporal resolution of LOAT can be optimized by increasing the absorbing particle density within the field of view, it will still remain inferior to the standard OAT where images can be acquired using single laser shots. However, considering that OAT can readily provide a higher spatio-temporal resolution in three dimensions as compared to other bio-imaging modalities \cite{dean2017advanced}, it may be still possible to image relatively fast biological events with LOAT. Localization of OA signals in the time domain may also be of interest. For example, it has recently been shown that OAT can monitor neuronal activity at otherwise unreachable spatio-temporal resolutions by detecting changes in the signals generated by genetically encoded calcium indicators \cite{dean2016functional}. By localizing brain activation events in both the spatial and temporal dimensions it may be possible to provide new insights into neuronal connectivity at scales and depths not accessible with fluorescence microscopy. The combined spatio-temporal information can also be used for tracking individual absorbers and measure the flow velocity.

In conclusion, LOAT can significantly enhance the well-established advantages of OA imaging by breaking the acoustic diffraction barrier at depths within the diffusive regime of light. It can also attain better visibility of vascular structures, thus improve the overall image quality in limited-view tomographic acquisition scenarios. Given the intrinsic sensitivity of optoacoustics to multiple hemodynamic and oxygenation parameters, LOAT may therefore enable new level of performance in studying functional and anatomical alterations of microcirculation.

\acknowledgements{Financial support is acknowledged from the European Research Council Grant ERC-2015-CoG-682379, National Institute of Health grant R21-EY026382-01, Human Frontier Science Program (HFSP) Grant RGY0070/2016 and the German Research Foundation Grant RA1848/5-1.}


\end{document}